**Final Report**

Age-at-harvest models as monitoring and harvest management tools for Wisconsin carnivores

Federal Aid in Wildlife Restoration, WI W-160-R

By Maximilian L. Allen, Nathan M. Roberts, and Timothy R. Van Deelen

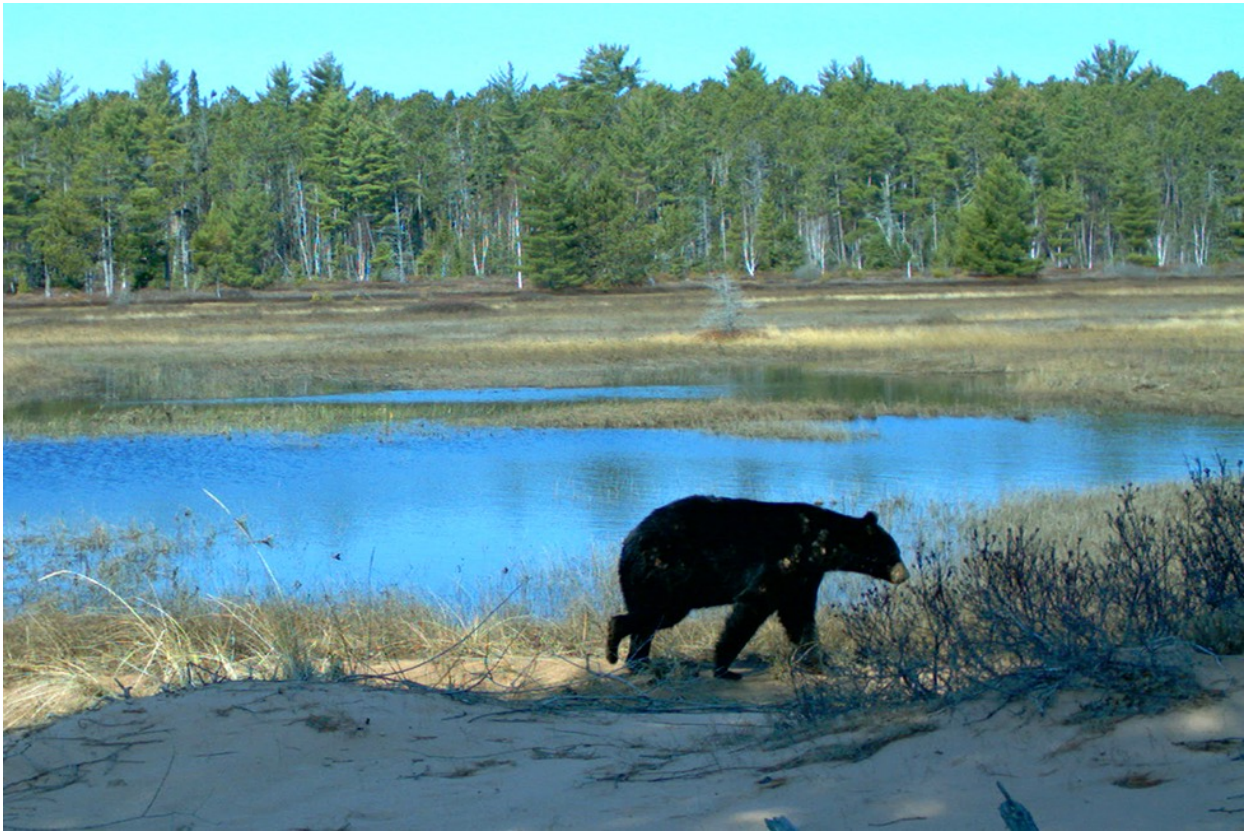

**Acknowledgements**

This is the final report for Federal Aid in Wildlife Restoration Grant WI W-160-R, with additional support from the Wisconsin Department of Natural Resources and the University of Wisconsin - Madison Department of Forest and Wildlife Ecology. Many people supported this project, including from the Wisconsin Department of Natural Resources (Scott Hull, David MacFarland, Robert Rolley, Brian Dhuey, Jess Rees, Catherine Dennison, Nicholas Forman, and Taylor Peltier), the University of Wisconsin – Madison (Lucas Olson, Qing Li, Yanshi Luo), Minnesota Department of Natural Resources (Andrew Norton), and University of Wisconsin – Stevens Point (Shawn Crimmins); we thank them for their support.

**Rationale for Project**

Quantifying and estimating wildlife population sizes is a foundation of wildlife management. However, many carnivore species are cryptic (Gese 2001; Allen et al. 2016), leading to innate difficulties in estimating their populations (Gese 2001; Hiller et al. 2011). Given the difficulties in enumerating carnivore populations, wildlife managers often rely on demographic data collected from harvested animals to estimate population parameters such as survival, recruitment, and population growth (e.g., Skalski et al. 2005). In Wisconsin, managers at the Department of Natural Resources (WDNR) use accounting style analyses to estimate wildlife populations. The drawback of these models is their lack of scientific rigor (i.e., they provide no estimates of error or variance), and over time they can lose track of the population without independent estimates. For example, the accounting style estimate underestimated Wisconsin's black bear (*Ursus americanus*) population by 2/3 compared to an independent estimate from tetracycline marking (MacFarland 2009), and there are similar concerns for the current furbearer population estimates.

Over the past decade there have been advances in many statistical models, including rigorous statistical models that use age-at-harvest (AAH) data and integrate auxiliary data (usually other harvest or demographic data) to accurately estimate populations. Models using AAH data are often most practical for hunted populations, especially when working with a population across large spatial scales when other methods of collecting data are difficult and costly (Skalski et al. 2005; Norton 2015). We distinguish between these models based on whether the method uses frequentist or Bayesian statistics and how the AAH data are modeled. Models based on frequentist statistics include statistical population reconstruction (SPR; e.g., Skalski et al. 2011) that integrated hunter effort, and state-space models such as Fieberg et al. (2010) that integrated food availability. Models based on Bayesian statistics include data augmentation models such as Conn et al. (2008) that is similar to SPR and integrated tag recovery, and state-space models such as Norton (2015) that integrated known-fate survival data.

In Wisconsin, evaluations of these approaches are needed to determine the long-term population dynamics of carnivore populations, and to determine if harvest-related demographic data should continue to be collected by wildlife managers. We evaluated the potential for using more rigorous statistical models to estimate the populations of black bears, and their applicability to other furbearers such as bobcats (*Lynx rufus*). Our first step was to collect and evaluate data. Harvest numbers themselves often tell an incomplete story, and many factors affect the number of animals harvested, so we also evaluated factors that can drive changes in demography and harvest rates over time. One gap in the WDNR data was auxiliary data (such as independent survival estimates), and to date there has not been a model developed that creates accurate estimates without integrating auxiliary data. Bayesian state-space models may be able to accomplish this, as one of their main strengths is appropriately using regularization to share information across space and time in the model, and efficiently using all available data compared to other modeling approaches.

## Evaluating Data and Creating Databases

The use of state-space models is possible with data collected by WDNR because, in most cases, it has been collected over multiple decades. The initial phase of our project was obtaining and evaluating data collected by the WDNR, and then creating master databases (Table 1). It is important for databases, not only to store the collected data, but also seamlessly integrate into the analyses the data is collected for.

Our method of setting up the data can be used as a template for furbearer data or other harvested species such as white-tailed deer (*Odocoileus virginianus*).

**Table 1.** Final black bear databases delivered to WDNR.

| Name | Years | Description |
|---|---|---|
| Bear Harvest | 1986-2017 | Summary of black bear harvest for state and zones by year and sex. |
| Bear County Harvest | 1971-2015 | Annual and 5-year mean harvest at the county level. |
| Bear Capture | 1979-2016 | Individual capture and harvest histories for 2,021 problem black bears. |
| Bear Hunter Surveys | 1994-2014 | Responses to periodic surveys of black bear hunters. |
| Bear Demographics | 1986-2016 | Individual demographic and harvest data for 67,984 harvested black bears. |
| Bear Reproduction | 1995-2014 | Reproductive histories for 4,688 harvested female black bears. |
| Bobcat Harvest | 1980-2014 | Summary of bobcat harvest, permits, and pelt price. |
| Bobcat Demographics | 1983-2015 | Individual demographic and harvest data for 6,202 harvested bobcats. |
| Bobcat Hunter Surveys | 2007-2015 | Responses to annual surveys of bobcat hunters. |

## Black Bear Research

Black bears are a spatially dispersed solitary carnivore (Rogers 1987; Taylor et al. 2015) that are K-selected (e.g., long-lived with heavy investment in parental care; Pianka 1970). Black bears are widely distributed across North America, with many populations expanding in recent years (Garshelis et al. 2016), and their ecology varies across their range (Beston 2011; Garshelis et al. 2016). In Wisconsin, black bears are a widespread game animal whose populations have been growing over the last 45 years (MacFarland 2009; Sadeghpour and Ginnett 2011). The number of black bears harvested over that time has also steadily increased (Figure 1). Most black bears in Wisconsin are found in the northern half of the state, but the population has been expanding southward in recent years.

Since 1985, the WDNR has estimated black bear populations using a deterministic accounting model (MacFarland 2009). However, independent capture-recapture estimates generated from tetracycline marking found that the current model underestimated the population size by nearly 2/3 (MacFarland 2009). This occurred in part because of the failure of the model to account for the functional relationship of bait stations and population size (MacFarland and Van Deelen 2011), and possibly also the inability of the deterministic model to account for variation in harvest and population demographics over time.

Independent population estimates have allowed the WDNR to more accurately assess the black bear population in the state (MacFarland 2009), but these are expensive and often conducted years apart.

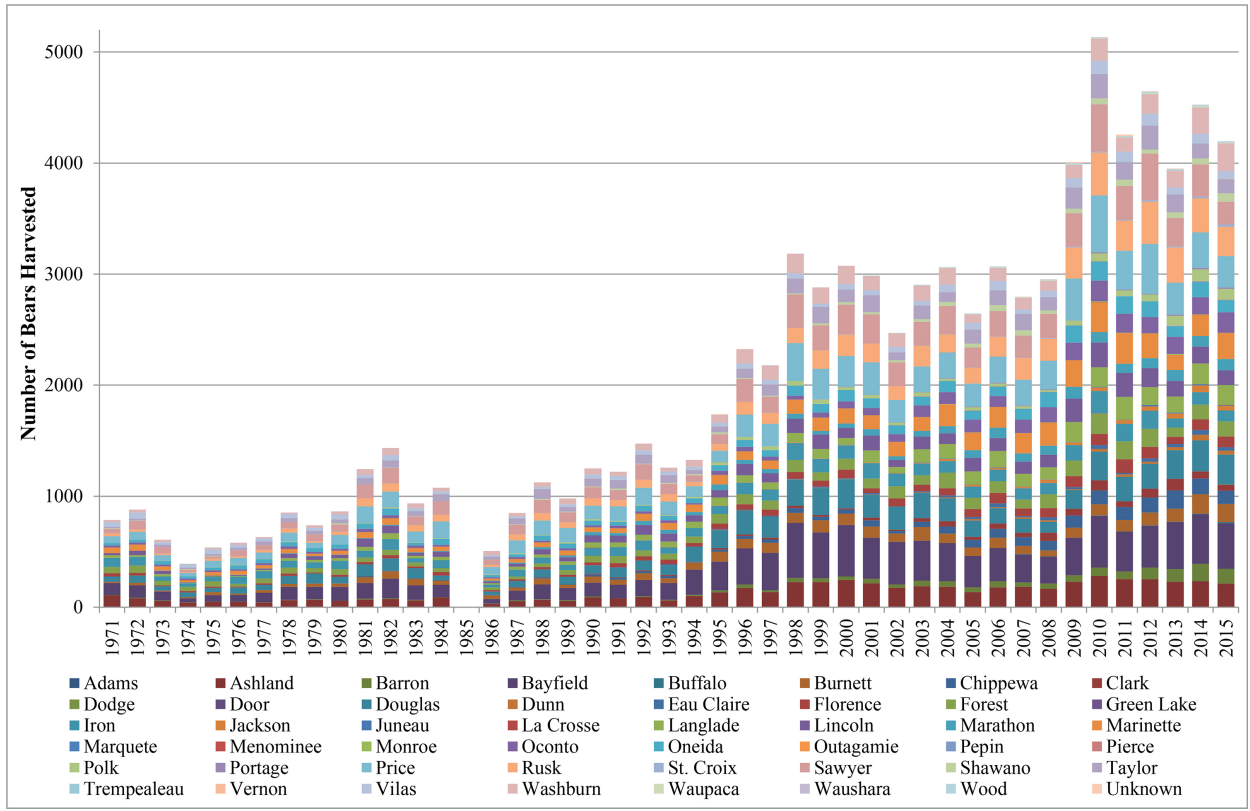

**Figure 1.** The number of harvested black bears in Wisconsin from 1971-2015, with no bear harvest in 1985. The number of harvested bears in each county is noted by a different color.

To assess whether an AAH state-space model would work for black bears we performed initial analyses of trend in the sex and age of black bears in the annual harvest. The age structure of harvested animals is essential to most population estimates based on harvest, including most integrated population models. We first grouped black bears into age classes, and then determined the annual proportion of harvest for each age class for each sex (Figure 2). We found that the age structure of both males and females harvested have been generally stable over time, even when substantial changes in harvest occurred. This suggests that the sex and age structure of harvested black bears is representative of the population, a key assumption of these models, and, thus the AAH data could be used as the basis of an updated statistical model.

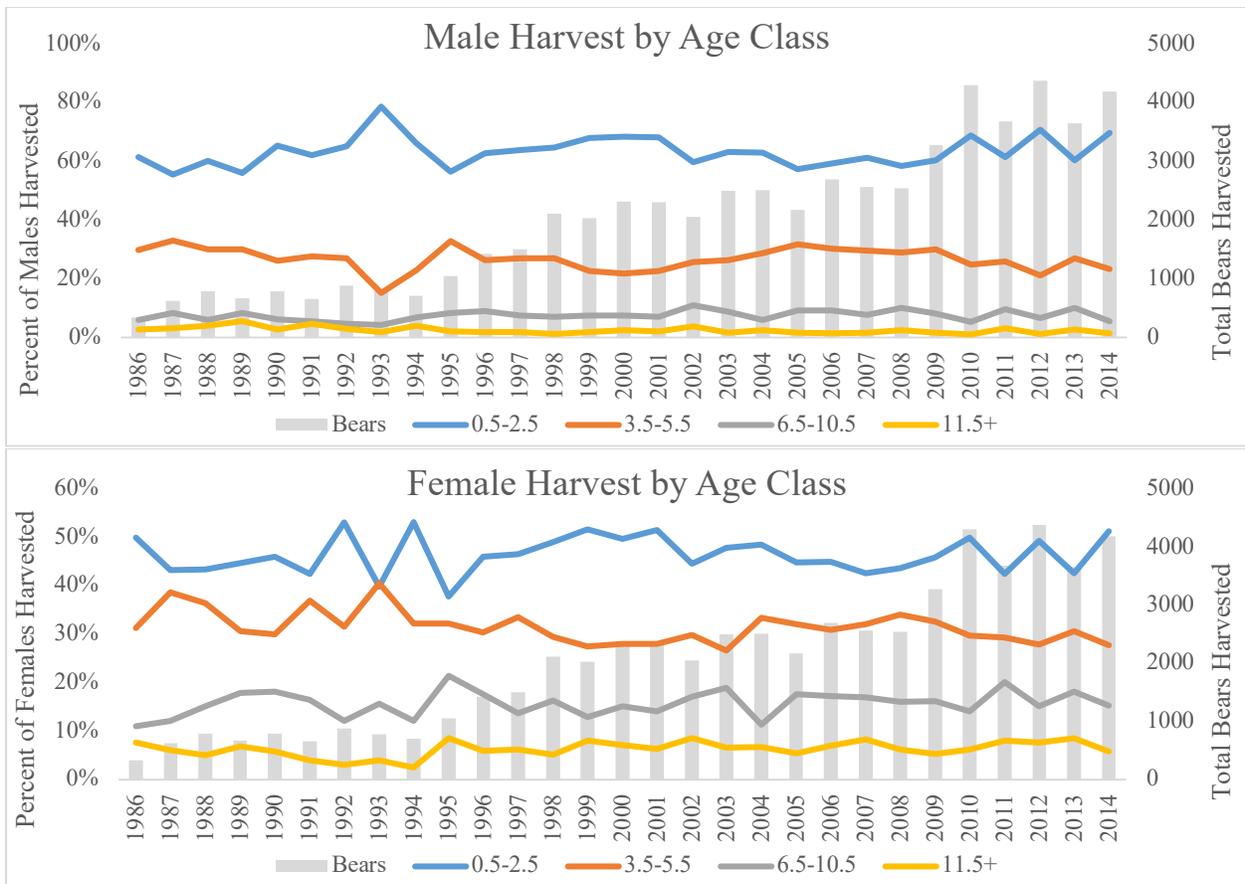

**Figure 2.** The pooled age structure of male (top) and female (bottom) black bears harvested in Wisconsin, 1986-2014.

We next evaluated the reproduction data that the WDNR has been paying Matson's Lab to estimate by analyzing successful litters for the cementum annuli of teeth from harvested black bears (Allen et al. 2017). Patterns of cementum annuli width may be able to indicate years with successful litters in black bears, and we evaluated reproduction estimates from cementum annuli of 19,101 black bears collected over 20 years in Wisconsin to determine the benefits and drawbacks of cementum-based estimates for management agencies. Unfortunately, the technique only worked for 25% of submitted samples, and 49% of estimates contained uncertain litters. Whether uncertain litters were counted or not caused significant variation in estimates of age at first litter, number of litters per female and interbirth intervals. Hence, naive treatment of uncertain litters may bias analyses. A dataset optimized to reduce bias suggested that litters per female ranged from 0 to 12, and having more litters was correlated with older age. Black bears in Wisconsin lived up to 31 years and were documented producing cubs into their late 20's. Mean successful interbirth interval was 2.07 years and increased as females aged. Large samples of teeth collected from harvested black bears over multiple decades potentially provides a wealth of information on reproductive parameters at a minimal cost compared to intensive field studies. The technique, however, does not estimate litter size, which makes it difficult to estimate potential long-term changes to litter size, and their effects on the population. Until uncertain litters are understood mechanistically and can be better quantified reproductive estimates from cementum annuli techniques should be interpreted with caution, and it is questionable whether the cost is worth the data received.

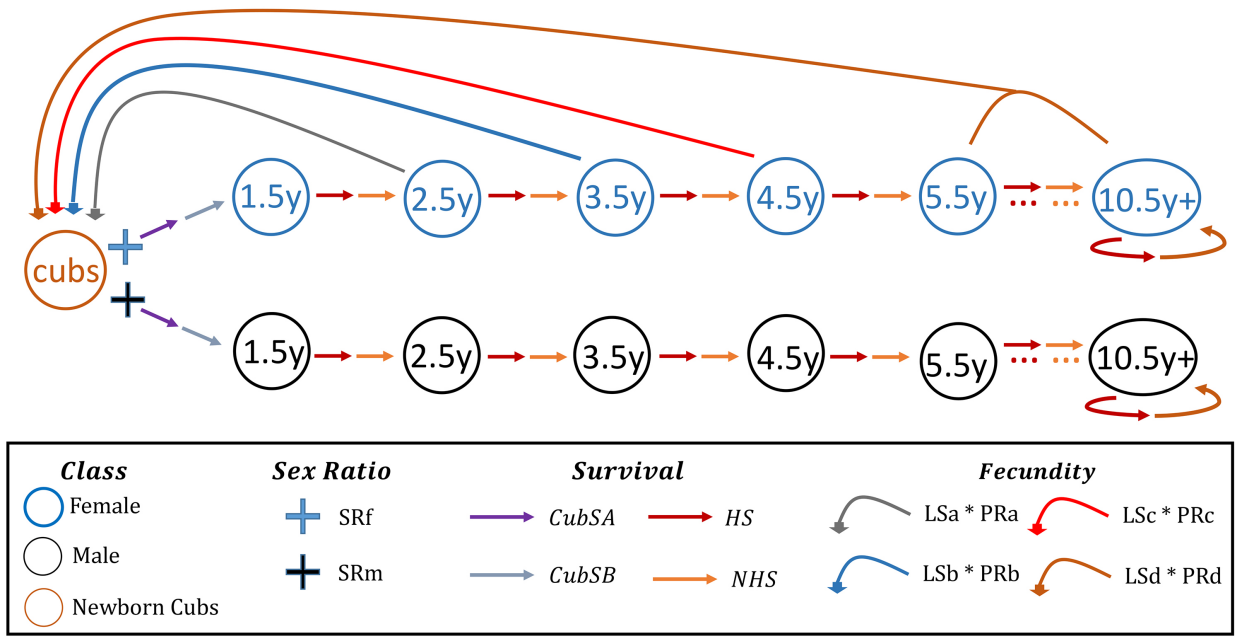

**Figure 3**. Life cycle diagram of black bears used to construct the stage-structured population matrix.

To estimate black bear populations, we developed an AAH state-space model in a Bayesian framework based on Norton (2015) that can account for variation in harvest and population demographics over time (Allen et al. 2018*a*). Our state-space model consists of two process models (population and observation) whose likelihoods were jointly modeled (Buckland et al. 2004; Norton 2015). The population process model was based on the unobserved/latent population matrix, and the observation state process was based on observed harvest data (Newman et al. 2009; Norton 2015). Our population process model was constructed as a two-sex, ten-stage population projection matrix (Caswell 2001) (Figure 3). The goal of our model was to estimate the total abundance (*N*) of the black bear population in each management zone in Wisconsin immediately prior to the hunting season. We used black bear harvest data collected from mandatory registration in Wisconsin to determine the total annual observed harvest (*O*), and the number of harvested animals with known age and sex (*C*).

**Table 1**. Descriptions of demographic parameters and their prior means and distributions.

| Variable | Parameter | Recruitment Priors Mean | Distribution | |
|---|---|---|---|---|
| *LS-a* | Litter Size 2.5-year-olds | 2.00 | Gamma (20,10) | |
| *LS-b* | Litter Size 3.5-year-olds | 2.00 | Gamma (20,10) | |
| *LS-c* | Litter Size 4.5-year-olds | 2.00 | Gamma (20,10) | |
| *LS-d* | Litter Size 5.5+ year-olds | 2.74 | Gamma (16.4,6) | |
| *PR-a* | Pregnancy Rate 2.5-year-olds | 0.003 | Beta (2.61,1000) | |
| *PR-b* | Pregnancy Rate 3.5-year-olds | 0.25 | Beta (34,100) | |
| *PR-c* | Pregnancy Rate 4.5-year-olds | 0.53 | Beta (54,48) | |
| *PR-d* | Pregnancy Rate 5.5+ year-olds | 0.48 | Beta (47,50) | |
| *SP* | Sex Proportion (female) | 0.46 | Beta (426, 500) | |
| | | Survival Priors | | |
| Variable | Parameter | Mean | Long-Term Precision | Annual Precision |
| *HSm* | Male Harvest Survival | 0.77 | 3 | Gamma (20,0.5) |
| *HSf* | Female Harvest Survival | 0.85 | 3 | Gamma (20,0.5) |
| *NS* | Non-harvest Survival | 0.95 | 4 | Gamma (20,0.5) |
| *CubSa* | Cub Survival years 0.0-0.5 | 0.84 | 4 | n/a |
| *CubSb* | Cub Survival years 0.5-1.5 | 0.71 | 4 | n/a |
| *Rep* | Recovery Rate | 0.98 | 2 | n/a |

We used reasonably informative prior distributions for the demographic parameters of the model (Table 1). Because demographic information for prior distributions from Wisconsin were sparse, these values were derived from a literature review of areas with similar habitats in North America (Allen et al. 2018*a*). Demographic parameters included litter size (*LS*), pregnancy rate (*PR*), cub sex ratio (*SR*), survival during harvest season (*HS*), survival during non-harvest season (*NS*), cub survival from birth to age 0.5 (*CubSa*), cub survival from age 0.5 to 1.5 (*CubSb*) and reporting rate (*Rep*). In the model we allow the survival values to vary by year, and this amount of variation is dictated by the long-term precision value, while the annual precision is the distribution that survival is sampled from.

We first created a state-wide estimate as a proof of concept for the model (full model information and findings in Allen et al. 2018*a*). The abundance estimate from our model was reasonably similar to an independent capture-recapture estimate from tetracycline sampling in 2011 (Figure 4), which we consider strong proof of concept for the model. In addition, the model was statistically robust to bias in the prior distributions for all parameters except for reporting rate. Deterministic models tend to be very sensitive to initial population size (Grund and Woolf 2004), but our model was also robust to bias in these estimates, and may be a key strength of this model when considered for use by management agencies. Our state-space model created an accurate estimate of the black bear population in Wisconsin based on age-at-harvest data and improves on previous models by using little demographic data, no auxiliary data, and not being sensitive to initial population size. Considering our findings, we consider this a valid method to improve the precision of population estimation for black bears in Wisconsin.

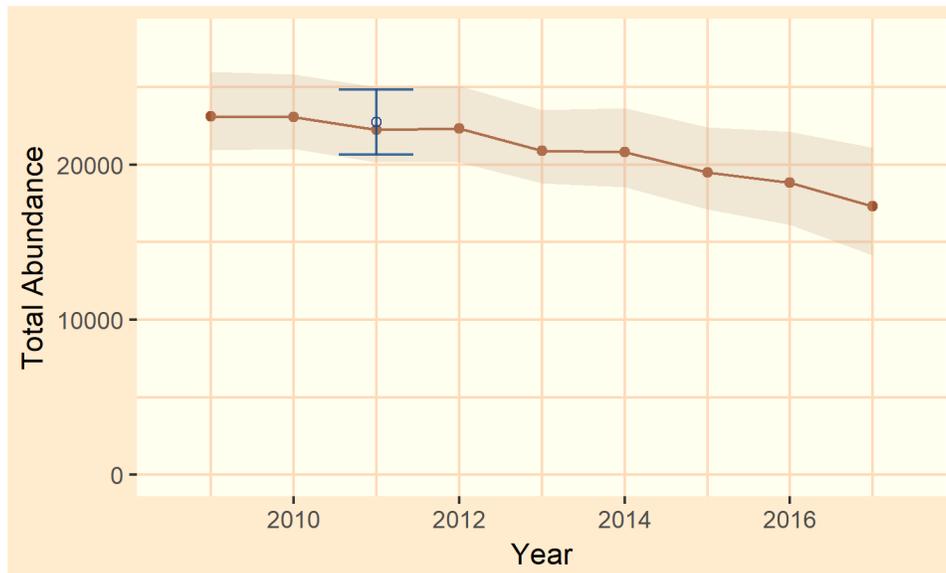

**Figure 4**. Statewide population estimates from an AAH state-space model for Wisconsin (2009 to 2017) in brown, with shaded area indicating 95% credible intervals. The independent capture-recapture population estimate (for bears 1.5+) from tetracycline marking in 2011, and 95% confidence intervals, are shown in blue.

We next evaluated the use of the model to create accurate zone-specific estimates. In the zone-specific models we based our initial adult population sizes for each zone ($N_A$ = 6,009, $N_B$ = 4,468, $N_C$ =

4,841, and $N_D$ = 6425) on independent capture-recapture estimate generated from tetracycline marking from 2011 (Rolley and Macfarland 2014). To account for known underestimates of females and overestimates of males in initial population estimates in the proof of concept model, most likely based on a slight bias in selection for male black bears (Malcolm and Van Deelen 2010), we considered 55% of the overall harvest in each age class to be the female proportion and 45% to be the male proportion.

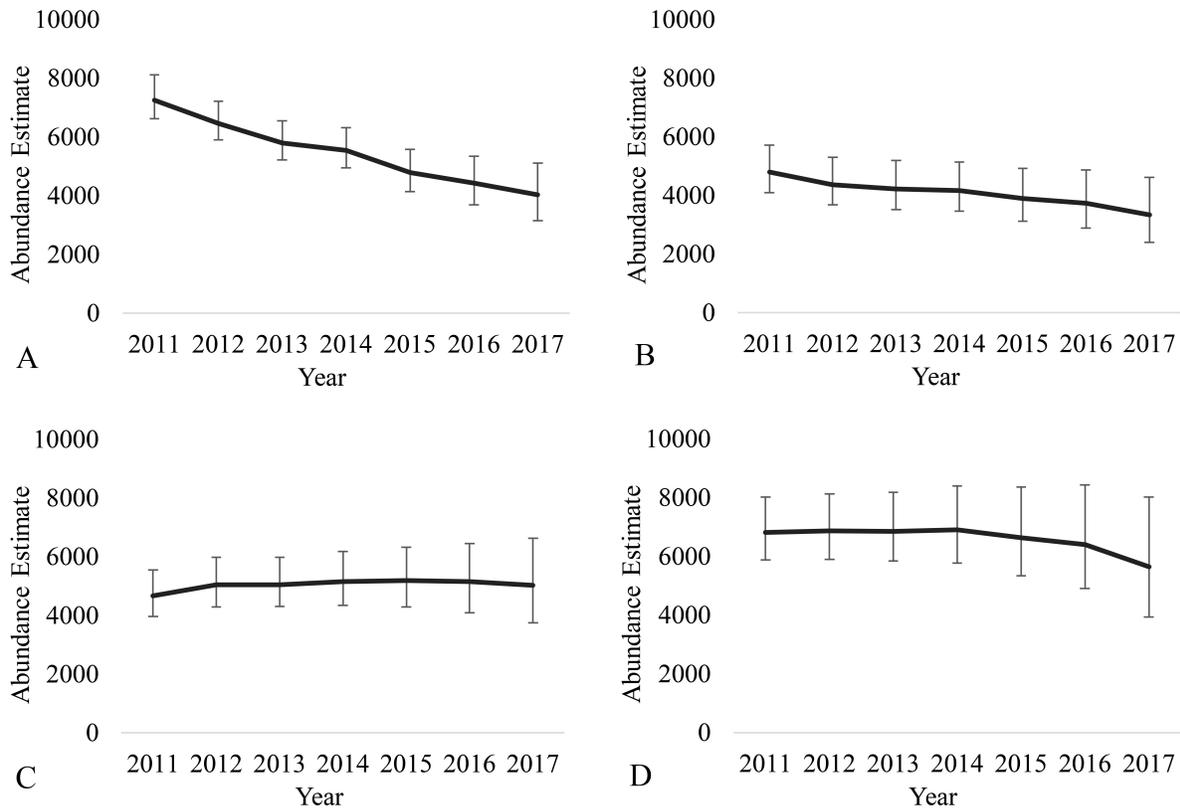

**Figure 5.** Results of AAH state-space model population estimation for each black bear management zone in Wisconsin and their 95% credible intervals.

The population estimates and 95% credible intervals from our AAH state-space models were reasonable for each management zone (Figure 5), and are likely more accurate than the accounting style estimates currently used by the WDNR (e.g., Macfarland 2009). Our population estimates show decreasing trends in zone A and relatively stable trends in the three other management zones. After the initial round of independent population estimates from tetracycline showed that the population had been underestimated by 2/3 (MacFarland 2009), the harvest quotas increased substantially beginning in 2009 and the number of black bears harvested annually in Wisconsin are now the highest of any state in the lower 48 states. The populations in each zone also showed slightly declining ages of harvested black bears, but these do not appear to be substantially affecting the population trend to date, although continued monitoring is warranted.

Overall, the AAH state-space models in a Bayesian framework appear effective for making zone-specific abundance estimates for black bears in Wisconsin. This statistically rigorous method improves upon the accounting style statistical method currently used by WDNR by incorporating annual variance and providing variance estimates for the population size and demographic parameters. Other auxiliary data can be integrated into the model to improve the accuracy of the estimates, however, including annual independent population estimates or observation surveys, survival estimates or other demographic parameters, or other covariates that affect demographic parameters such as winter severity or snow depth. State-space models are flexible, and can be adjusted to any harvest system, including those with unique data or parameters. The increased accuracy of the AAH state-space models should allow for better management by being able to set accurate quotas to ensure a sustainable harvest for the black bear population in each zone.

## Implementing the Model

The full model code for the statewide population estimate is in Appendix A (A1 'Run_Model' is the code to run the model, A2 'JAGS_Model' is the model itself, and A3 'Figures' is the code to create figures). Run_Model is the only file that needs to be adjusted before running a new set of models, unless you are making substantial changes to the model itself. Using the model assumes basic knowledge of Program R (R Core Team 2017), and a working knowledge of JAGS (Plummer 2003) would be beneficial. In this example, we will assume that the modeler wants to add a new year of harvest data and then run the statewide model. The modeler would a) adjust the number of years, b) consider adjusting the starting population values, c) provide harvest data (total harvest, and sex and age of harvested bears), and d) adjust prior values as they see fit.

The first step is to adjust the number of years (Y) on line 36 from 8 to 9. If the modeler would like to adjust the starting population they would adjust the values for females on line 42 and for males on line 43. The modeler would then provide the observed harvest data as separate csv files without headers for rows and columns.

For example, the observed harvest data (saved as a file named O) can be set up as

| Year | Female | Male |
|------|--------|------|
| 2009 | 1851 | 2158 |
| 2010 | 2608 | 2525 |
| 2011 | 2066 | 2191 |
| 2012 | 2317 | 2329 |
| 2013 | 1917 | 2035 |
| 2014 | 2240 | 2286 |
| 2015 | 2000 | 2198 |
| 2016 | 2371 | 2311 |
| 2017 | X | Y |

where X and Y are the values for the new year. This is then formatted to be read by the model by removing the headers, as

| | |
|------|------|
| 1851 | 2158 |
| 2608 | 2525 |
| 2066 | 2191 |
| 2317 | 2329 |
| 1917 | 2035 |
| 2240 | 2286 |
| 2000 | 2198 |
| 2371 | 2311 |
| X | Y |

and saved as O.csv.

The ages of harvested females is set as A1.csv and the ages of harvested males is set up as A2.csv. The data is set up (in this case the female data) as

| | | | | | | | | | | |
|------|-----|-----|-----|-----|-----|-----|----|----|----|-----|
| 1507 | 370 | 316 | 261 | 127 | 107 | 72 | 74 | 36 | 36 | 108 |
| 2163 | 532 | 549 | 304 | 209 | 129 | 101 | 85 | 42 | 41 | 171 |
| 1773 | 401 | 349 | 226 | 155 | 139 | 96 | 95 | 54 | 69 | 189 |
| 2139 | 491 | 555 | 287 | 184 | 134 | 88 | 83 | 62 | 53 | 202 |
| 1744 | 406 | 334 | 281 | 150 | 105 | 73 | 86 | 56 | 55 | 198 |
| 2048 | 550 | 535 | 256 | 196 | 120 | 100 | 68 | 67 | 46 | 110 |
| 1649 | 391 | 364 | 262 | 155 | 101 | 90 | 54 | 40 | 29 | 163 |
| 1941 | 542 | 430 | 276 | 177 | 130 | 96 | 54 | 40 | 41 | 155 |

where each column is an age class (1.5 years old to 10.5+ years old), and each row is a year with the most recent year at the bottom. The demographic parameters are found from lines 66 to 93, the truncation values from lines 96 to 108, and the precision from lines 111 to 119. The modeler can adjust the prior distribution by adjusting the appropriate values in this section.

## Bobcat Research

Bobcats are a widely distributed, and territorial, solitary felid (Bailey 1974; Larivière and Walton 1997; Allen et al. 2015). Bobcats were likely distributed throughout Wisconsin before European colonization (Klepinger et al. 1979), and a bounty system for bobcats was used in the state from 1867 through 1964 (Rolley et al. 2001). Mandatory registration of bobcat harvest began in 1973 (Rolley et al. 2001), and in 1980 the WDNR implemented a bag limit of 1 per license coincident with a hunt that was restricted to the northern 1/3 of the state (Creed and Ashbrenner 1975; Rolley et al. 2001). The current quota system which limits tags to 1 per season and is based on preference points (hunters who are unsuccessful in drawing a tag are awarded "preference points" that weight their application in the licence lottery and increase the probability of drawing a tag during subsequent drawings) was implemented in 1992 (Rolley et al. 2001), and a zone for the southern 2/3 of the Wisconsin was added in 2014. Various factors have contributed to these changes in the management of bobcats in Wisconsin, resulting in a decreasing supply of tags and an increasing bobcat population. Over the last two decades, along with the decreasing supply of available tags, hunter participation and success rate has increased; a trend that is contrary to the general stable or declining trend in hunter interest across North America (Miller and Vaske 2003; Winkler and Warnke 2013). We used annual harvest data and surveys to determine the factors responsible for bobcat harvest and hunter participation in Wisconsin.

The decrease in supply of tags was strongly correlated with increases in number of applications for tags and hunter success (percent of filled tags) (Figure 6). The increased interest in bobcat hunting (applications and participation) also correlated with each other and with success (Figure 7). The bobcat population abundance and number of tags issued was the common variable among the top models for annual bobcat harvest and hunter participation, highlighting the critical role supply of available tags has in furbearer management and hunter participation. The restricted bobcat tags and increasing success have occurred over time, and correlate with substantially increasing both hunter populations and hunter interest. Contrary to other furbearer research, pelt price and other socioeconomic factors had less importance in our models than management variables, potentially because the lower supply of tags transitioned the bobcat harvest to a trophy hunt.

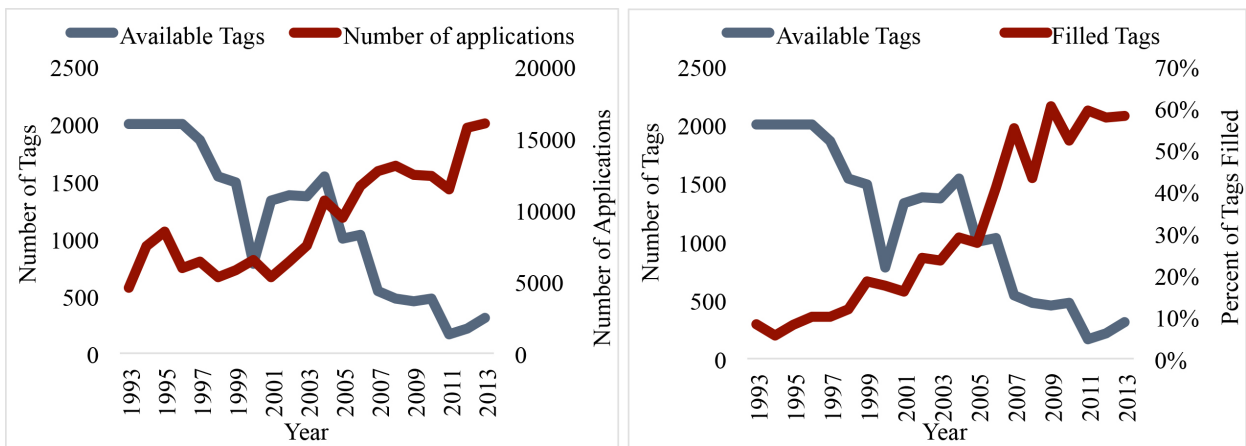

**Figure 6.** The inverse correlation of the number of tags issued (blue line) with the number of applications (orange line in left figure) and the percent of filled tags (orange line in right figure).

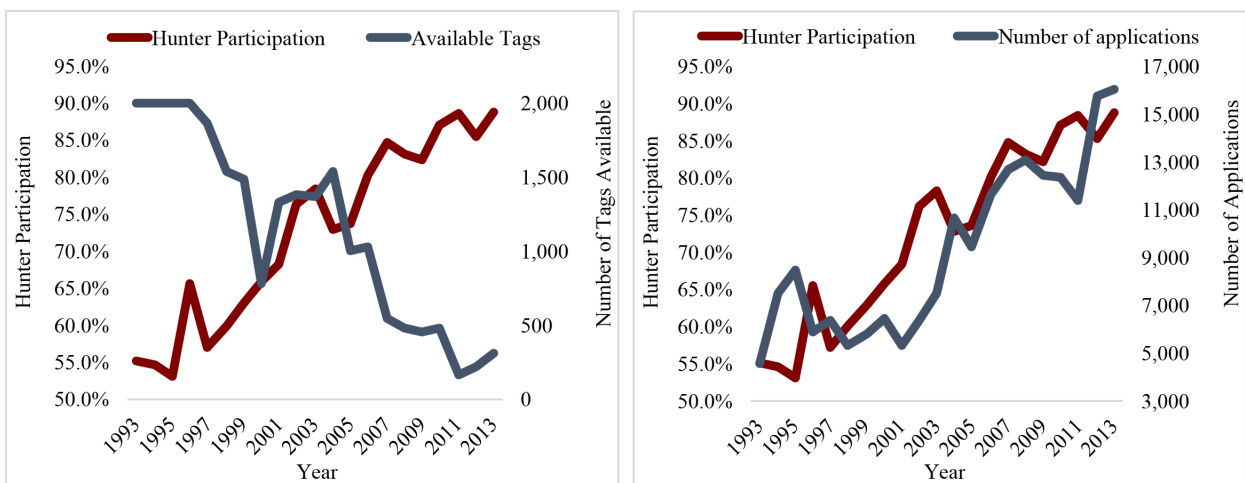

**Figure 7.** Annual hunter participation (red line) and the inverse correlation with number of tags available (blue line in left figure) and the correlation with number of applications (blue line in right figure).

Wildlife researchers often rely on demographic data collected from harvested animals to estimate population dynamics, as is the case with the AAH state-space model we developed for black bears. But harvest data may be non-representative if hunters/trappers have the ability and motivation to preferentially select for certain physical traits because harvest itself can also affect the sex and age structure of the population (Ginsberg and Milner-Gulland 1994; Langvatn and Loison 1999; Bischof et al. 2009; Pelletier et al. 2012). Male animals in populations are often selected for because of trophy traits (Martínez et al. 2005; Mysterud et al. 2006; Mysterud 2011; Pelletier et al. 2012), or are more vulnerable to harvest due to larger movement patterns and riskier behavior (McLellan et al. 1999; Bischof et al. 2009; Mysterud 2011). Bobcats can be harvested by trapping, calling (using auditory lures to bring the bobcat within range of the waiting hunter), or by hunters who use hounds to find, pursue, and tree them. The increases in perceived value of the harvested animal that we previously showed could potentially lead to increased selectivity for animals with specific traits (i.e. trophy animals) that could bias sampling of harvested animals from the population. Wisconsin's cumulative annual quotas of tags and harvested bobcats have been reduced over time compared to neighbouring states, and Kapfer and Potts (2012) suggested that decreasing tags in Wisconsin has increased the perceived value of an opportunity to participate in a hunt and disassociated bobcat hunting from the traditional value of being a source of income from pelts. We used data from bobcats harvested in Wisconsin (1983-2014) to understand if harvest method and demographics of harvested bobcats have changed over time, and if bobcat hunters/trappers exhibited selection (Allen et al. 2018*b*).

Each trait of harvested bobcats that we tested (mass, male:female sex ratio, and age) changed over time. These selected traits were interrelated, and we inferred that harvest selection for larger size biased harvests in favor of a) larger, b) older, and c) male bobcats by hound hunters. The selected traits of harvested bobcats were interrelated, as males are usually larger than females and size also increases with age, so we inferred that harvest selection for larger size led to selection of older male bobcats as has also been shown for other harvested felid species (Nilsen et al. 2012). We also found an increase in the proportion of bobcats that were harvested by hound hunting compared to trapping from 1973-2014, and that hound hunters were more likely to create taxidermy mounts from their harvested bobcats than trappers. This increased selectivity has led to substantial changes in the characteristics of harvested bobcats, with a preference for individuals with trophy traits (older age, and males over females) to make into taxidermy mounts, and makes bobcat hunting in Wisconsin more closely akin to trophy hunting.

      Selection by hunters may bias population models that are based on the demography of harvested bobcats, and accounting for biases that may occur, including from different harvest methods, is critical when using harvest-dependent data (Skalski et al. 2005), or that any bias is constant over time (Harris and Metzgar 1987; Skalski et al. 2005). Due to these factors, the AAH data is too biased to be accurate when used as the basis for a population model (whether an accounting style model or AAH state-space model). The most effective way to use this data would be with an independent population estimate as a starting point, and other auxiliary integrated into the model. It is possible that just data from trappers could be used, but this is a limited sample size due to the restricted harvests and overall decline in trapper numbers. For these reasons, developing an AAH state-space model for bobcats is currently not possible unless methods are developed to account for the bias. The AAH state-space models should work well, however, for other furbearers such as river otters (*Lontra canadensis*) and fishers (*Pekania pennanti*) which are primarily trapped.

**Future Work and Recommendations**

      Our AAH state space model was built specifically for black bears, but a similar model can be built for other harvested species that have a reasonable number of individuals in the harvest with known sex and age. Our model worked well because the WDNR has attempted to age and sex every harvested black bear, but the models also perform well when only a proportion of animals are aged (Norton 2015), assuming the sample size is large enough. Most state management agencies have collected sex and age data for harvested animals over the course of decades, and our model could therefore be widely applicable. A key strength of the model is its robustness to prior values, and other data can be integrated

into the model to increase the accuracy of the population estimate. The use of this model should allow of more accurate estimation of harvested wildlife species.

The use of AAH state-space models is possible with data collected by WDNR because in most cases it is long-term, having been collected over multiple decades. There is incredible value in these long-term datasets, but they need to be built in a way that allows them to integrate into the statistical models, including Bayesian-based population models, in order for the given software to read and analyze the data. For example, the data from all years needs to be merged into one cohesive file, rather than the WDNR system of creating independent files for each year with different files for each annual report. In addition, the different files often do not have an identifier that will allow the files from different databases to be merged with each other. For example, it is not possible to link the age of many harvested bears with the method of its harvest or county/management zone.

Although the AAH state-space model is robust and effective, it is not a magic bullet, and it needs to be calibrated periodically with independent estimates. The WDNR should consider having independent estimates of the population (as was previously done with tetracycline for bears), in order to incorporate as auxiliary estimates for the population models. Ideally these would be done every 3 years, but every 5 years would suffice. Tetracycline is currently in moratorium by the federal government, and may be permanently restricted. Other options to consider are capture-recapture techniques, such as spatially explicit capture recapture analyses (Royle et al. 2013), with either genetic or camera data (such as that collected by Snapshot Wisconsin).

A strength of our state-space model was its robustness to biased prior distributions, including initial population size. The Percent Relative Change values when changing prior means by 10% for all parameters were < 2.00%, except for reporting rate. This is encouraging, because we derived many of our prior distributions from literature values and several of the parameter estimates appear to be almost entirely dependent on the prior distributions. Lack of information about parameter values can cause problems in population estimation models (e.g., Skalski et al. 2005), especially when models are sensitive to parameters that are determined by expert opinion that can itself be biased (Hristienko et al. 2007).

These informed values help the model perform better than completely uninformed parameters, although the model could potentially be improved by using data that are specific to the study area. Considering how robust the state-space model is to biased prior distributions, and the applicability of using prior distributions informed by the literature review, the priority for future work should focus on accurately determining the reporting rates. Data collection, in the form of surveys, have now been implemented based on our recommendations to determine accurate reporting rate values for Wisconsin.

At this time, we consider the WDNR model used to estimate bobcat abundance to be ineffective and inaccurate. We found that bobcats harvested by hunters were more likely male, older, and larger, which appears to be due to selection of bobcats with trophy traits. This AAH data is too biased to be effective in many population models unless accounted for. In this case, bobcats harvested by trappers may be more representative of the population; although trappers may exhibit selection for bobcats with certain traits as well. If trappers were non-selective, then it would be possible to use the demography of bobcats harvested by trappers for population estimates. Trapping has been declining in Wisconsin, however, and the trapped harvest may not yield sufficient sample sizes. It may also be possible to account for the bias statistically, but that was beyond the scope and timeline of this project. As a result, the WDNR implemented a bobcat working group in 2018 to explore these problems. Solutions decided upon by this group include using other metrics such as catch-per-unit-effort and harvest success to help monitor the population.